\documentclass[12pt]{article}
\usepackage{amsmath}
\usepackage{epstopdf}
\usepackage{mathrsfs}
\usepackage{natbib}
\usepackage{dcolumn}
\usepackage{graphicx}

\def\be{\begin{equation}}
\def\ee{\end{equation}}
\def\bea{\begin{eqnarray}}
\def\eea{\end{eqnarray}}
\def\nn{\nonumber}

\makeatletter \@addtoreset{equation}{section}

\begin{document}
\begin{titlepage}
\vspace*{1mm}%
\hfill%
\vbox{
    \halign{#\hfil        \cr
           IPM/P-2009/040 \cr
                     } 
      }  
\vspace*{15mm}%
\begin{center}

{{\Large {\bf  Weyl Anomaly in NonRelativistic CFTs }}}

\vspace*{15mm} \vspace*{1mm} { Ali Davody}

 \vspace*{1cm}
{\it  School of physics, Institute for Research in Fundamental Sciences (IPM)\\
P.O. Box 19395-5531, Tehran, Iran \\ }

\vspace*{.4cm}

{\it  Department of Physics, Sharif University of Technology \\
P.O. Box 11365-9161, Tehran, Iran}

\vspace*{.4cm} {  davody@ipm.ir}

\vspace*{2cm}
\end{center}

\begin{abstract}
We study  Weyl symmetry  for non-relativistic conformal filed
theories  on curved spatial  spaces, and calculate it's quantum
anomaly. We show that there is no \emph{geometric anomaly}, and
the  non-relativistic Weyl anomaly can appear only due to
interaction. Also we study the  anomaly by using  the light-cone
approach.

\end{abstract}

\vspace{7cm}
\begin{center}
\it{Dedicated to Amir Abbass Varshovi}
\end{center}

\end{titlepage}

\section{Introduction}
Recently non-relativistic version of AdS/CFT correspondence  has
received a lot of interest(a partial list of them is\cite{barobach}\footnote{ It was also  shown
that this non-relativistic correspondence can be generalize to the fermionic systems \cite{Akhavan:2009ns}.}). Non-relativistic CFTs (NRCFTs),
are invariant under Schr\"{o}dinger group, which contains galilean
group as a subgroup\cite{Hagen:1972pd,Mehen}. Actually the
Schr\"{o}dinger group has two elements more than the galilean
group: non-relativistic scale  and special conformal
transformation. Under  nonrelativistic-scaling  (NR-scale)  time
and space scale differently

\be\label{nrscale}
 t\longrightarrow e^{\sigma} t \;\;\;,\;\;\; x\longrightarrow
e^{\frac{\sigma}{2}}x
\ee and the special conformal transformation is given by
\be
t\longrightarrow\frac{t}{1+\alpha t} \;\;\,\;\;x\longrightarrow\frac{x}{1+\alpha t}
\ee

These transformations must combine  with an appropriate change of fields for  the  action to  be invariant.
It is also  known (see  \cite{Goldberger:2008vg}  and references therein)
that the Schr\"{o}dinger algebra can be derived  from  relativistic conformal algebra
in one higher dimension. To see this,  let us  consider  a  complex
massless scalar field in $d+1$ dimensions which is invariant under conformal group $SO(d+1,2)$

\be
S=\frac{1}{2} \int d^{d+1}x \partial _\mu \Phi\partial^\mu \Phi^\dag.
\ee
By going to the light-cone coordinate

\be
t=\frac{x^0+x^d}{\sqrt2} \;\;,\;\; \xi=\frac{x^0-x^d}{\sqrt2},
\ee
and taking $\Phi$ to have a definite momentum in $\xi$ direction, $\Phi=e^{iM\xi}\Psi$, we arrive at

\be\label{flatfreeaction}
S=\int dt d^{d-1}x  (-i M\Psi^\dag\partial_t\Psi+i M\Psi
\partial_t\Psi^\dag+\partial_i\Psi\partial^i\Psi^\dag)
\ee
which is the  free Schr\"{o}dinger action in $(d-1)+1$ dimensions and is invariant under Schr\"{o}dinger
group. Therefore the  Schr\"{o}dinger group may be thought of as a subgroup of conformal group in one higher dimension
that does not mix modes with different momentum $M$ along the null direction $\xi$ \cite{Goldberger:2008vg}.

An important quantity in field theories  is energy-momentum tensor, which  must
be traceless in  Weyl invariant theories on a curved space.
 Actually this symmetry is anomalous  and one interesting feature of $AdS/CFT$ correspondence
 is that the Weyl anomaly of boundary field theory can be computed from gravity dual \cite{Henningson:1998gx}.
In analogy with relativistic case, the energy-momentum tensor of
NRCFTs must obey the  trace identity, and one may expect that by using  the holographic renormalization for
non-relativistic backgrounds (see\cite{Taylor:2008tg} ) an anomaly appears in the Ward identity of energy-momentum
tensor. So it is interesting  to consider  NRCFTs on curved special
space and study the issue of Weyl anomaly for these  theories.


Let us review some aspects of the CFTs on curved spaces \cite{Duff:1993wm}.
Indeed it is well known that \emph{Weyl}$\otimes$\emph{diff}  invariant  theories are invariant
under the  conformal group  of the background.
  To see this, consider  the Weyl transformation

 \be
 G_{\mu\nu}\longrightarrow e^{\sigma} G_{\mu \nu},
 \ee
 and note also  that under the   conformal  transformations   the metric changes
 by a factor which  can be absorbed  by a Weyl transformation. Motivated by this definition
 of Weyl transformation, we will  define the NR-Weyl symmetry which  guaranties
  that the theory becomes  invariant under NR-scale transformation  (\ref{nrscale}) on flat
space.

As in the relativistic case where scale invariance implies that the
improved energy-momentum tensor is traceless \cite{Callan:1970ze}

\be
\Theta^\mu_\mu=0,
\ee
the NR-scale symmetry  implies that the \textit{non-relativistic trace} of
improved energy-momentum tensor is zero too

\be\label{trace}
2T^{00}-T^i_i=0.
\ee
On the other hand it is known that relativistic Weyl symmetry breaks at  the quantum level \cite{Duff:1993wm}
\be
\Theta^\mu_\mu={\cal{A}}(G_{\mu\nu}),
\ee
where the anomaly ${\cal{A}} $  depends on the geometry of space. A
natural question is whether the equation  (\ref{trace}) holds at the  quantum
level too\footnote{see \cite{Bergman:1991hf} for calculation of scale anomaly
in a non-relativistic system  on flat space}? Actually we find that
NRCFTs do not admit \emph{Geometric Anomaly}, i.e. NR-Weyl anomaly is
zero for free NRCFTs on curved spatial space, though   NR-Weyl
anomaly can appear in the interacting theories. This is different   from the
relativistic case where the free CFTs admit Weyl anomaly as well.

The organization of this paper is as  follows. In the next section we give a definition of NR-Weyl symmetry.
In section 3  we calculate the NR-Weyl anomaly. In section 4  we analyze anomaly from
light-cone point of view. We end in section 5  with conclusions and some comments.

\section{Weyl Symmetry in NRCFTs}
When a relativistic field theory on curved space
 has Weyl symmetry, it becomes a CFT on flat space. In a similar way we can define NR-Weyl symmetry for
NR-field theory on curved spacial space in such  way that a NR-Weyl
invariant filed theory becomes NRCFT on the flat space\footnote{Here by
NRCFT we mean a galilean  field theory which is invariant under NR-scale
 transformation (\ref{nrscale}), though  in all cases that we will
consider the special conformal transformation is also a symmetry of
action in flat space }. In other words NR-scale symmetry must be in the residual
symmetries  of NR-Weyl invariant field theory when the metric
becomes flat. Now consider the following transformation

\be\label{NRW}
h'_{ij}=e^\sigma h_{ij}   \;\;\;\ , \;\;\;\ t'=e^\sigma t
\ee
where $h_{ij}$ is the metric of special space and $\sigma$ is a constant.
Suppose  a field theory  is invariant under the above transformation,
it is easy to see  that it becomes a NRCFT in flat space. Actually under
NR-scaling (\ref{nrscale}), the time and flat metric transform as
$ t\longrightarrow e^\sigma t \;\;  , \;\;  \delta_{ij}\longrightarrow e^\sigma \delta_{ij}$
 and this factors can be absorbed by a NR-Weyl transformation such that the action remain invariant.
 So  symmetry under (\ref{NRW})  guaranties  the field theory  becomes invariant under
 non-relativistic scale transformation  (\ref{nrscale}) in flat space. Hence we  take the (\ref{NRW})
  for the definition of NR-Weyl  transformation.

To  proceed let us  consider the simplest case; free Schrodinger field in
2+1 dimensions  on a curved spatial space

\be\label{freeaction}
S_{free}=\int dt d^2x \sqrt h {\cal{L}}=\int dt d^2x \sqrt h (i M\Psi^\dag\partial_t\Psi-i M
\partial_t\Psi^\dag\Psi-h^{ij}\partial_j\Psi^\dag\partial_i\Psi),
\ee
where $M$ is mass of the particle. This action is invariant under NR-Weyl  transformation  (\ref{NRW}) if $\Psi$
transforms in the following way:

\be\label{Wpesi}
\Psi^{'}(x,t')=e^{-\frac{\sigma}{2}}\Psi(x,t)
\ee
NR-Weyl  symmetry  implies that\footnote{here we use relativistic notation but only raising
and lowering of spatial index is meaningfull.}

\be\label{noether}
\delta h_{ij}\frac{\partial(\sqrt h\cal{L})}{\partial h_{ij}}+
\partial_\mu[\frac{\partial( \sqrt h\cal{L})}{\partial (\partial_\mu\Psi )} \delta\Psi+
\frac{(\partial \sqrt h\cal{L})}{\partial (\partial_\mu\Psi^\dag )} \delta\Psi^\dag+ \sqrt h {\cal{L}} \delta x^\mu ] = 0.
\ee
The first term is spatial part of canonical  energy-momentum tensor
\be\label{emtf}
T^{ij}_f=\frac{2}{\sqrt h}\frac{\delta S}{\delta h_{ij}}=h^{ij}(i M \Psi^\dag\partial_t\Psi-i M
\partial_t\Psi^\dag\Psi-\partial^k\Psi^\dag\partial_k\Psi)+\partial^j\Psi^\dag \partial^i\Psi+\partial^i\Psi^\dag \partial^j\Psi.
\ee
By making  use of  this expression, the equation  (\ref{noether}) can be recast to the following form

\be\label{nrtrace}
\frac{1}{2}h_{ij}T^{ij}-iM\Psi^\dag\partial_t\Psi+iM\partial_t\Psi^\dag\Psi=0.
\ee
One could   also add interacting  terms to the action which   preserve
the NR-Weyl symmetry. For example the  following interactions are NR-Weyl
invariant:

\bea\label{interactions}\nn \int  d^2x \sqrt h
\frac{g}{4}\Psi^\dag\Psi^\dag\Psi\Psi
\;\;,\;\;
 \int  d^2x \sqrt h d^2y \sqrt h \Psi^\dag
(x)\Psi^\dag(x) \frac{1}{|x-y|^2}\Psi(y)\Psi(y) \eea
the first one describes Non-relativistic bosons interacting via a $\delta$ function
 potential with strength $g$ \cite{Bergman:1991hf}, while  the second one describes non-relativistic
 particles interacting through a $\frac{1}{r^2}$ potential. We  now  examine the validity
 of  classical identity  (\ref{nrtrace}) at  the quantum level.


%

\section{NR-Weyl Anomaly}
Quantum anomaly comes from the renormalization of the theory. Actually the matrix elements of energy-momentum tensor are
divergent and must be regularized  before taking trace in
(\ref{nrtrace}), and it is not clear whether  the  regularized  energy-momentum
tensor obeys  NR-trace identity (\ref{nrtrace}). Let us first
consider the free field  (\ref{freeaction}), which  can be expanded in
terms of energy eigenstates, $\psi_n's$:


%
\be\label{solution}
\Psi(x,t)=c \sum_n a_n  \psi_n(x) e^{-i\omega_n t}
\ee

%

According to canonical quantization we have (by appropriate choice of
$c$ in (\ref{solution}))

\be
[a_n,a_m^\dag]=\delta_{n,m}.
\ee
The vacuum  state is  defined  by $a_n|0\rangle=0$, and the exited states can be built by acting on $|0\rangle $
with the creation operators. Since there is no negative energy in non-relativistic case the mode expansion
(\ref{solution}) does not contain creation operator, $a^\dag$,  the expectation value of
 free energy-momentum tensor (\ref{emtf}) between any states is finite and so NR-trace identity (\ref{nrtrace})
holds at the  quantum level for free NRCFTs\footnote{Note that we have written
 the action and energy-momentum tensor in the normal ordering form.}. In other words unlike relativistic case,
  there is no \emph{Geometric anomaly} in NRCFT's.

On the other hand  for interacting theories, classical symmetry can be broken due to quantum corrections. Indeed
NR-Weyl transformation (\ref{NRW}) includes  time scaling and  thus  the scale of energy is changed
 by this transformation. If $\beta$ function is nonzero, this implies that the shift
  of the coupling constant, $g$,  under NR-Weyl transformation is
\be
g\rightarrow g-\sigma \beta(g),
\ee
and the Lagrangian is changed  in the following way
\be
{\cal{L}}\rightarrow{\cal{L}}-\sigma \beta(g)\frac{\partial {\cal{L}}}{\partial g},
\ee
So that the  classical identity (\ref{nrtrace}) will be corrected by the quantum correction as follows \cite{Peskin:1995ev}:
\be\label{nrwanomaly}
2T^{tt}-T^i_i={\cal{A}}=-\frac{\beta}{\sqrt h} \frac{\partial {\cal{L}} }{\partial g }.
\ee
Therefore the  NR-Weyl anomaly is given by  $\beta$ function which
depends on the details of interaction  and the geometry of space. To  clarify how the geometry of space
enters  the  calculating of $\beta $ function, we will compute the $\beta $ function for
an interacting bosonic system on a sphere  in the next section.

\subsection{  Anyons  on sphere}
Consider a system of bosonic particles  in $2+1$ dimensions interacting via a $\delta$-function
 potential on a sphere. The action of this system can be derived by taking NR limit from the relativistic
$\lambda \phi^4$ theory, and is given by (we set $M=1$ )
\be\label{iaction}
S=\int dt d^2x \sqrt h (i \Psi^\dag\partial_t\Psi-i \Psi
\partial_t\Psi^\dag-h^{ij}\partial_i\Psi\partial_j\Psi^\dag+\frac{g}{4}\Psi^\dag\Psi^\dag\Psi\Psi)
\ee

The  free field solution on sphere can be expanded in terms
of spherical harmonic functions $Y_{l,m}(\theta,\phi)$:

\be \Psi(x,t)=\frac{1}{  R}\sum_{l,m} a_{l,m}Y_{l,m}(\theta , \phi) e^{-i E_l t}
\ee
where $E_l=\frac{l(l+1)}{2R^2}$ and $R$ is radius of the sphere. The Feynman propagator in position
space is



\bea\label{feynmanpropagator} D_F(x,t,x',t')&=&\langle0|T \Psi(x,t)
\Psi^\dag(x',t')|0\rangle =\frac{1}{ R^2}\theta(t-t')\sum_{l,m} Y_{l,m}(\theta
,\phi) Y^*_{l,m}(\theta' ,\phi')e^{-i E_l( t-t')}\nn\\&=&\frac{1}{R^2}\sum_{l,m}\int\frac{ d\omega}{2\pi
i}\frac{e^{-i\omega(t-t')}}{\omega-E_l+i\epsilon} Y_{l,m}(\theta
,\phi) Y^*_{l,m}(\theta' ,\phi') \eea
while  in the  momentum space it is given by
\bea\begin{split}
D_F(l,m,\omega,l',m',\omega')&=\int d^2x \sqrt h
 d^2x' \sqrt h dt   dt' e^{i( \omega t-\omega' t')}Y^*_{l,m}(x) Y_{l',m'}(x')
\langle0|T \Psi(x,t) \Psi^\dag(x',t')|0\rangle\cr
&=R^2(2\pi)\delta(\omega-\omega') \delta_{l,l'}\delta_{m,m'} \frac{1}{\omega-E_l}\
\end{split}
\eea

%

\begin{figure}
 \includegraphics[scale=.5]{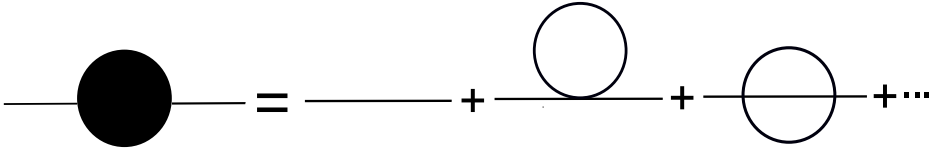}\caption{two-point function}\label{2pointfunction}
 \end{figure}

 Now consider the two-point function(see Fig\ref{2pointfunction})

\be \langle\Omega|T
\Psi\Psi^\dag|\Omega\rangle=\langle0| T \Psi\Psi^\dag
\exp(-i\int dt d^2x\sqrt h H_I)|0\rangle_{connected}\ee
 due to $\theta$ function in (\ref{feynmanpropagator})
and normal ordering in action (\ref{iaction}) all loop contributions are zero.
So the mass and field are  not renormalized.

\begin{figure}
 \includegraphics[scale=.5]{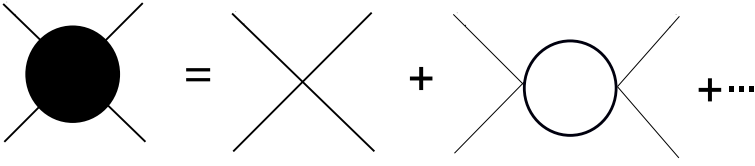}\caption {four-point function}\label{4pointcomplet}
 \end{figure}

To evaluate  the     $\beta$  function, we must compute
 the four-point function at  one-loop level by inserting  a \emph{renormalization condition} at scale $\mu$,
  which we assume   to be

\bea\label{rc}
\widetilde{G}(l_1,l_2,l_3,l_4,m_1,m_2,m_3,m_4)
\bigg{|}_{\begin{split}l_2=l_4=0 , l_1=l_3=\mu\;\;\;\; \cr m_1=m_2=m_3=m_4=0 \end{split}}=-ig
\eea
where $\widetilde{G}$  is the four-point function in the  momentum space
\footnote{We drop the energy conserving $\delta$ function
and normalization due to integration over $Y_{l,m}$s ,and also external propagators }.
 Up to one-loop order we have (see Fig\ref{4pointcomplet})

\bea\nonumber
\widetilde{G}(\mu)=-ig -\frac{ig}{2 R^2} \sum_{l_5,l_6}\int \frac{d\omega_5}{2\pi i}
\frac{1}{(\omega_5-E_{l_5})(\mu-\omega_5-E_{l_6})} \int d^2z
\sqrt h Y^*_{l_1,0}(z)Y^*_{0,0}(z) Y_{l_5,m_5}(z)Y_{l_6,m_6}(z)
\\ \nn
\int d^2w \sqrt h Y_{l_1,0}(w)Y_{0,0}(w)
Y^*_{l_5,m_5}(w)Y^*_{l_6,m_6}(w)
\eea
\bea\nn
=-ig -\frac{ig^2}{8\pi  R^2}\sum^{\Lambda}_{l_5,l_6}
  \frac{1}{E_{l_5}+E_{l_6}-\mu}
\frac{(2l_5+1)(2l_6+1)}{(2l+1)}(l_5,l_6,0,0|l,0)^2\;\;\;\;\;\;\;\;\;\;
\eea
where $\Lambda$ is an ultraviolet cut off. Renormalization condition
(\ref{rc}) implies that the counterterm, $\delta g$, must be


\be\label{counterterm}
\delta g =\frac{ g^2}{8\pi MR^2}\sum^{\Lambda}_{l_5,l_6}
\frac{1}{E_{l_5}+E_{l_6}-\mu^2}
\frac{(2l_5+1)(2l_6+1)}{(2 l+1)}(l_5,l_6,0,0|l,0)^2.
\ee
leading to a non zero  $\beta$ function defining by  $\beta(g)=\mu\frac{\partial g }{\partial\mu}$, and the
   NR-Weyl anomaly is given by:
\be\label{nrwanomalysphere}
2T^{tt}-T^i_i=-\frac{\beta(g)}{4} \Psi^\dag\Psi^\dag\Psi\Psi
\ee

It is  interesting to study   $R\longrightarrow\infty$ limit of NR-Weyl anomaly.
 To get a meaningful result  in $R\longrightarrow\infty$  limit, we must take angular momentum $\l$
to be large. Also we approximate the sum in (\ref{counterterm}) by integration and calculate
the integral for large $\l_5$. By using  the asymptotic expression for Clebsh-Gordon coefficients \cite{Asymptotics}

\be
 (l_5,l_6,0,0|l,0)^2\approx \frac{2}{\pi l_5\sin\theta}
\ee
with  $\theta$ is the  angel between $\l$ and $\l_5$, the equation  (\ref{counterterm}) becomes:

\be\begin{split}
\delta g=\frac{ g^2}{8\pi }\int \frac{d\theta dl_5   }{l_5-l \cos\theta}\cr
=\frac{g^2}{8\pi}\ln\frac{4\Lambda^2}{\mu^2}  .
\end{split}
\ee
The  anomaly can be read from (\ref{nrwanomalysphere}) as follows

\be\label{flatanomaly}
{\cal{A}}=\frac{g^2}{16\pi } \Psi^\dag \Psi^\dag \Psi\Psi
\ee
in agreement with the result of \cite{Bergman:1991hf}.

\section{Light-Cone Weyl Anomaly}
As noted  in the introduction, NRCFTs  can be derived from CFTs in
one higher dimension by light-cone procedure. In this section we
generalize  this  approach to the curved space. Since CFTs on curved
space are Weyl invariant, let us start with the
  \textit{Weyl}$\otimes $\textit{diff}-invariant  complex massless scalar filed  in 3+1 dimensions
\be\label{s}
S_{1+3}=\frac{1}{2}\int d^4x \sqrt G (G^{\mu\nu}\partial_
\mu\Phi\partial_\nu\Phi^\dag-\frac{1}{6} R \Phi\Phi^\dag)
\ee
This action is invariant under relativistic Weyl transformation
\be
G^{\mu,\nu}\longrightarrow e^{\sigma(x)} G^{\mu,\nu} \;\;\;\;,\;\;\;\;
\Phi(x)\longrightarrow e^{\frac{-\sigma(x)}{2}}\Phi(x)
\ee
which implies that the energy-momentum tensor is traceless

\be\label{rtrace}
\Theta^\mu_\mu=0,
\ee
This  identity receives  quantum anomaly

\be\label{ranomaly}
\Theta^\mu_\mu={\cal{A}}(G_{\mu\nu})
\ee
which  depends on the geometry of space-time.

To study   NR-field theory on curved spatial space we choose the  following form for the metric

\be\label{fixedmetric}
ds^2=-dx^2_0+dx^2_3+h_{ij}dx^idx^j.
\ee
Going to the light cone coordinates

\be\label{lcmetric}
ds^2=-2dtd\xi+h_{ij}dx^idx^j
\ee
and assuming  that the   $\Phi$ has a definite light-cone momentum $\Phi=e^{iM\xi}\Psi$,
the equation (\ref{s}) becomes

\be\label{action2}
S_{1+2}=\int dt d^2x \sqrt h (i M\Psi^\dag\partial_t\Psi-i M\Psi\partial_t\Psi^\dag
+h^{ij}\partial_i\Psi\partial_j\Psi^\dag-\frac{1}{6} R_h \Psi\Psi^\dag).
\ee
From the equation  (\ref{rtrace}) or by using Noether current, one can show
that the  energy-momentum tensor of above action, $T_{\mu\nu}$, satisfies  the following equation

\be
2T^{tt}-T^i _i=0.
\ee
If the light-cone procedure works
in quantum level, the equation  (\ref{ranomaly}) should reduce to

\be\label{lcanomaly}
2T^{tt}-T^i_i={\cal{A}}(h_{ij}).
\ee
On the other hand   we saw in section 2 that NR-Weyl anomaly occurs only in interacting
theories. Actually by the same approach of section 2 one can show that
NR-Weyl anomaly is zero for (\ref{action2}) and indeed the equation (\ref{lcanomaly})
 is not correct. Thus the light-cone procedure dose not lead to the correct result at the  quantum level.

Therefore a natural question would be what is the  symmetry of the action (\ref{action2})?
Since we have started from a \textit{Weyl}$\otimes$\textit{diff}-invariant
action and then the metric has been  fixed in the form of
(\ref{fixedmetric}) and also  we have kept a sector with a  definite light-cone momentum,
 the residual symmetry of the equation  (\ref{action2})  consists  of  those transformations, ${\cal O}$,
 that satisfy the following conditions
\be\label{symmetry}
{\cal O}\in CKV(\widetilde{G})   \;\;\;\;\;\;and\;\;\;\;\;\;[{\cal O},P_\xi]=0
\ee
where $CKV(\widetilde{G})$ is the set of conformal Killing vectors of the metric (\ref{lcmetric}).

In order for  the action    (\ref{action2}) to have  nontrivial
symmetry \footnote{time translation is the symmetry of
(\ref{action2}) for every $h_{ij}$},   the metric $h_{ij}$ must have
CKV. If we write (\ref{action2}) on a compact 2 dimensional
manifold, we have only two choices, $S^2$ with 6 CKV and $T^2$ whit
2 CKV. The conformal Killing vectors of  (\ref{lcmetric}) which
satisfy (\ref{symmetry}) for $S^2$ are

\bea\begin{split}
&  H=\partial _t \;\;,\;\;P_\xi =\partial_\xi\cr
&L_1=\sin\phi\partial_\theta+\cot\theta\cos\phi \partial_\phi
\;\;\;,\;\;\ L_2=\cos\phi\partial_\theta+\cot\theta\sin\phi
\partial_\phi \;\;\;,\;\;\ L_3=\partial_\phi
\end{split}
\eea
 so the symmetry algebra  of (\ref{action2}) on sphere is $SU(2)\times U(1)\times U(1)$. In
other words, the action is only invariant under the isometry of sphere, not by those C.K.Vs
which give a factor to  the metric  of sphere. Also we have $U(1)\times U(1)\times U(1)$ symmetry for torus.

\section{Discussions}

With  definition  of  Weyl transformation in  NR field theories, we have
shown that NR-Weyl anomaly is not a geometric effect and only appears in
interacting theories. An important difference between the Weyl
symmetry which  we have defined in (\ref{NRW}) and  those in the relativistic case  is that
the relativistic Weyl transformation is local, while the NR-Weyl
transformation is not local  in the sense that $\sigma$ in (\ref{NRW})
is not a function of position. Actually the free action is not
invariant under local NRWeyl transformation. It would be interesting
 to write down  an action with local NR-Weyl symmetry and study the theory on flat space
 with conformal symmetry in special space\cite{work}( see  \cite{Hosseiny:2009jj} for  affine extension
a class of nonrelativistic algebras including non centrally-extended
Schrodinger algebra and Galilean Conformal Algebra (GCA) in 2+1
dimensions).

We have also discussed anomaly from  light-cone point of view, where we have observed
 that the light cone procedure dose not work  at the  quantum level
%
 and it would be interesting to explore the
exact relation between NRCFTs and CFTs in one higher dimension at
the  quantum level \cite{work}.

It would be also very interesting to extend the results of this
paper  to Lifshitz-like theories which do not admit galilean symmetry but have
anisotropic scaling symmetry

\be t\longrightarrow \lambda^z  t \;\;\;\,\;\;\; x^i\longrightarrow
x^i\ee
For example for  the action

\be
S=\int dt d^dx  (\dot{\phi}^2 -  (\triangle\phi)^z)
\ee
due to the  presence of  the \emph{negative energy }, one would
expect that such theories admit \emph{geometric anomaly}. (see \cite{Adam:2009gq} for calculation of
Weyl anomaly for a four-dimensional z=3 Lifshitz scalar coupled to Horava's theory of anisotropic gravity )

\subsection*{Acknowledgements}
The author would like to thank Farhad Ardalan and Mohsen Alishahiha  for many helpful discussions and encouragement
 and for carefully reading and commenting on the manuscript.
 I would also like to thank Hamid R.Afshar, Amin Akhavan, Davod Allahbakhsi, Reza Fareghbal,Hanif Hadipur,
  Ali Naseh, Ali Vahedi  for discussions. I also like to thank Amir E. Mosaffa  for many useful
  discussion about different aspects of Weyl symmetry and Weyl anomaly and reading the manuscript.

\end{document}